\documentclass[a4paper,fleqn,usenatbib]{mnras}

\usepackage{txfonts}
\usepackage[T1]{fontenc}
\usepackage{ae,aecompl}
\usepackage{graphicx}	
\usepackage{amsmath}	
\usepackage{amssymb}	
\usepackage{wasysym}

\usepackage[T1]{fontenc}
\usepackage{aecompl}

\newcommand{\acb}{$\alpha$~Cen~B}

\newcommand{\logr}{$\log{R'_{HK}}$}
\newcommand{\fei}{\ion{Fe}{i} $4383$~\AA~}
\newcommand{\feii}{\ion{Fe}{i} $4404$~\AA~}
\newcommand{\feiii}{\ion{Fe}{i} $4375$~\AA~}

\title[Chromospheric Emission: $\alpha$ Cen B]{The Changing Face of $\alpha$~Centauri~B: Probing plage and stellar activity in K-dwarfs.}
\author[A. P. G. Thompson et al.]{
A. P. G. Thompson$^{1}$ \thanks{E-mail: athompson1501@qub.ac.uk},
C. A. Watson$^{1}$,
E. J. W. de Mooij$^{1,2}$ and 
D.  B. Jess$^{1}$
\\$^{1}$Astrophysics Research Centre, School of Mathematics and Physics, Queen's University Belfast, BT7 1NN, Belfast, UK
\\$^{2}$School of Physical Sciences, Dublin City University, Glasnevin, Dublin 9, Ireland\\}

\date{Accepted XXX. Received YYY; in original form ZZZ}

\pubyear{2016}

\begin{document}
\label{firstpage}
\pagerange{\pageref{firstpage}--\pageref{lastpage}}
\maketitle

\begin{abstract}
A detailed knowledge of stellar activity is crucial for understanding
stellar dynamos, as well as pushing exoplanet radial-velocity detection
limits towards Earth analogue confirmation. We directly compare archival
HARPS spectra taken at the minimum in \acb's activity cycle to a
high-activity state when clear rotational modulation of \logr~is visible.
Relative to the inactive spectra, we find a large number of narrow
pseudo-emission features in the active spectra with strengths that
are rotationally modulated.
These features most likely originate from plage, spots, or a
combination of both.
They also display radial velocity variations of $\sim$300 m s$^{-1}$ --
consistent with an active region rotating across the stellar surface.
Furthermore, we see evidence that some of the lines originating from
the `active immaculate' photosphere appear broader relative to the
`inactive immaculate' case.
This may be due to enhanced contributions of e.g. magnetic bright
points to these lines, which then causes additional line broadening. 
More detailed analysis may enable measurements of plage and
spot coverage using single spectra in the future.


\end{abstract}


\section{Introduction}
When trying to take precise radial velocity (RV) measurements of stars the
presence of activity contributes additional `jitter' to the RV signal that makes
exoplanet detection more difficult.
As such, the community makes use of the \logr~activity indicator to gauge the
detectability of planets and to better constrain RV jitter.
This measure, which traces changes of the cores of \ion{Ca}{ii} H \& K, was
first done by \cite{Wilson1978}, with the long baseline of measurements allowing
for activity cycles (similar to the $\sim$11-year cycle of the Sun) to be mapped for 
other stars  (e.g. \cite{Hall2007, Flores2016} and references therein).

\cite{Lovis2011} looked at stars observed with the High Accuracy Radial
velocity Planet Searcher (HARPS) instrument and found that $61\%$ of
the 304 FGK stars sampled show periodic variations.
They concluded that  activity cycles can induce RV variations having long period
and amplitude up to about $25$ ms$^{-1}$.
This result demonstrates the need to better understand the activity of exoplanet
host stars especially when searching for Earth analogs.

\cite{Dumusque2012} studied the RVs of \acb~(a $5,214\pm{33}$ K, K1V star) looking for the existence of a planet.
They used \logr~to get a better handle on the RV jitter, which shows a ramping
up of activity over the course of the observations.
The data from the most active nights display a clear periodic variation (see their Fig. 2)
caused by active regions rotating in and out of view.
Although not the main result from the paper, the \logr~values show a star going from
relatively quiet to active, which provides an interesting test bed to investigate the changes
that activity may have on the spectra of K-dwarfs. 

\begin{figure*}
	\vspace{-1.3cm}\includegraphics[width=\textwidth]{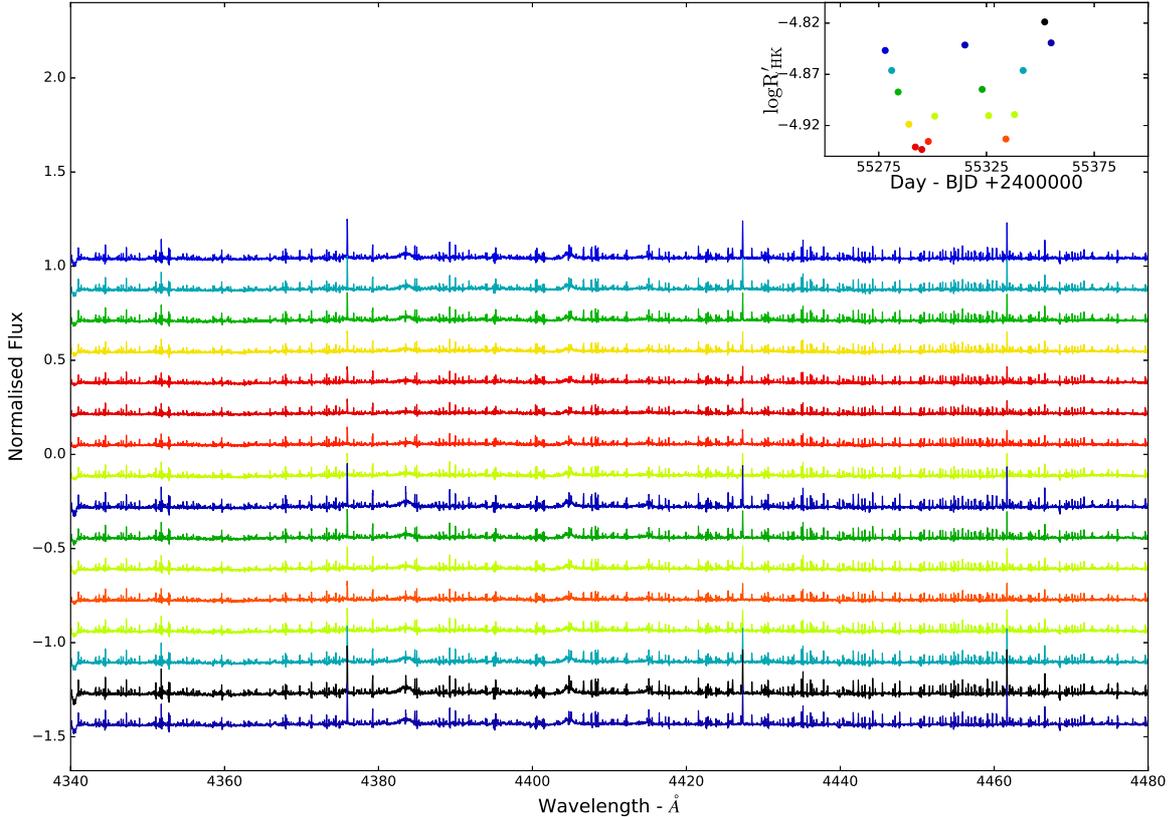}	
	\vspace{-1cm}\caption{A selection of relative spectra from the March-June 2010 period
	-- generated by dividing high-activity spectra by the master low-activity template
	-- for ease of viewing only 16 of the 48 weighted spectra available for this period are shown.
	The broad features seen at $4383$~\AA~and $4404$~\AA~correspond to
	temperature sensitive \ion{Fe}{i} lines.
	A large number of narrow `pseudo-emission' peaks can also be seen
	with the feature at $4375$~\AA~showing an excursion of $\sim{20\%}$.
	The colour of the residuals corresponds to the activity as seen in the \logr~
	(see insert at the top right).
	Note the change in the strength of all these features clearly correlates
	with the rotation cycle of the star.}
	\label{fig:temp_evolve}
\end{figure*}

While the \logr~of a star does give an indication of activity, the measure traces
changes in the chromosphere. 
We attempt to better constrain photospheric activity by investigating changes in
other spectral lines as a function of \logr~activity.
In this work we present results of comparing spectra taken during during high-
and low- activity phase of \acb.
In section \ref{lab:data} we discuss the data used in this analysis, with more
detail on the data processing given in section \ref{data processing}.
In section \ref{AandD} we discuss the changes observed in the generated `relative'
spectra and attempt to model the different morphologies of the observed narrow
pseudo-emission peaks.
In section \ref{linewidths_section} we measure the equivalent width of the 
pseudo-emission features and finally, in section \ref{lab:RVs}, we look at the radial velocity
shift that these peaks exhibit.

\section{Data}\label{lab:data}

We obtained archival HARPS data of \acb~from  February 2008 to July 2011
as used by \cite{Dumusque2012}.
This data covers a significant fraction of \acb's activity cycle,
spanning a range in \logr~from approximately -5 to -4.82.
The full dataset consists of 9693 spectra, though in this work we
focus our attention on data from 2010 March 23 to 2010 June 12,
hereafter referred to as the March-June 2010 period.
This range shows clear rotational variability in \logr~covering $\sim$2 of
\acb's $36.2$ day rotation periods \citep{DeWarf2010}.
This period is well sampled, with a total of 2475 spectra taken over 48 separate nights.

\subsection{Data Processing} \label{data processing}

The spectra were all aligned onto a common wavelength grid,
after correcting for the radial velocity (RV) shifts as published
by \cite{Dumusque2012} (including the orbital motion of the binary,
light contamination from $\alpha$ Centauri A, and the barycentric motion of the Earth).
The spectra were then all standardised to a common flux level.
This was done by dividing each individual spectrum by a high signal-to-noise
reference and fitting a 4th order polynomial to the resulting relative spectrum
over the spectral range of  $4300-5300$~\AA.
Each individual polynomial was then applied to its respective spectrum
in order to match their continuum to that of the reference.

For the purposes of this work, we were interested in the difference between
high- and low-activity spectra.
In order to do this, we identified the night with the lowest stellar activity measure
(as defined by \logr), which occurred on 28$^{th}$ February 2008.
We then stacked all of the data from this night to form a master low-activity
template spectrum, after following the process described earlier.
In this case we used the highest signal-to-noise spectrum from 28$^{th}$
February 2008 as the reference in the continuum matching process.
The end result was our master low-activity template.

For the 48 nights during the more active March-June 2010 period
we used this master low-activity template as the reference for the continuum matching. 
For each night we produce a single nightly spectrum by stacking the continuum
matched spectra using a weighted average.~
The root-mean-square of the relative spectrum (produced by dividing
the spectrum by the master low-activity template) was
measured in the spectral range $5050-5300$~\AA~and used as the weighting factor.
A rejection criterion of  greater than 1$\%$ in the measured root-mean-squared was
included to remove any wrongly labelled $\alpha$ Centauri B spectra
(some erroneous observations of $\alpha$ Centauri A took place) or
spectra obviously affected by echelle order mis-match.

\vspace{-0.9cm}\section{Analysis and Discussion} \label{AandD}

Once all of the nightly spectra were created using the process described
in section \ref{data processing} we generated `relative' spectra by dividing
each of the nightly outputs by the master low-activity template.
These relative spectra then highlight the differences between high- and low-activity
of \acb.

Visual inspection of the relative spectra for the March-June 2010 period was performed. 
A representative region ($4340-4480$~\AA) is shown in Fig. \ref{fig:temp_evolve}
and highlights the range of features we observe
(to help with observing the changes in the relative features only 16 of the 48
spectra available are shown in Fig. \ref{fig:temp_evolve}, these 16 are evenly spaced over the period to cover
the entire range of \logr).
Each relative spectrum has been colour coded with respect to its value of \logr~
with an equivalently coloured plot of \logr~ versus time shown at the top right of Fig. \ref{fig:temp_evolve}.

A number of broad features are seen in the relative spectra.
Some examples of these can be seen in Fig. \ref{fig:temp_evolve} at
$4383$~\AA~and  $4404$~\AA, and correspond to \ion{Fe}{i} species that are
used as spectral type indicators due to their temperature sensitivity \citep{Giridhar2010}.
The broad peak of the \ion{Fe}{i} lines indicate a change in temperature of the star.
As these lines are known to be photospheric in origin this temperature change
may indicate the presence of cooler active regions (i.e. spots) on the surface of \acb.

In contrast, numerous sharp `pseudo-emission' peaks can be seen,
the most prominent in Fig. \ref{fig:temp_evolve} occur at $4375$~\AA,
$4427$~\AA~and $4462$~\AA. The \feiii~line, for example, shows a peak at approximately
the $20\%$ level, which suggests a significant line change between
the high-activity spectrum compared to the master low-activity template.
This is similar to the findings of \cite{Basri1989}.
However, we note that \cite{Basri1989} constructed similar
relative spectra, but used {\em different} stars to represent high-
and low- activity cases.
This meant that their results were somewhat inconclusive, as
the authors could not be certain that the features they saw were
activity driven, or caused by differences in the metallicities, age,
$v\sin{i}$, temperature, surface gravity etc. between the active
and inactive stars -- a point raised by \cite{Basri1989} in their analysis.

Since we see morphologically similar results as reported by \cite{Basri1989},
(but without the confusion generated by using different stellar types in the analysis),
our work confirms that the bulk of the features reported by \cite{Basri1989} were
indeed likely to have been activity driven.
The fact that we also see these features modulated on the stellar rotation
period of \acb~further strengths this conclusion.
The features reported are not due to tellurics,
as these look distinctly different.

The strength of all the relative features change alongside the periodic
modulation of \logr.
We investigate this change further in section \ref{linewidths_section}
by measuring the pseudo-equivalent width of the features in the relative spectrum.

\begin{figure}
	\vspace{-0.8cm}\includegraphics[width=1\columnwidth]{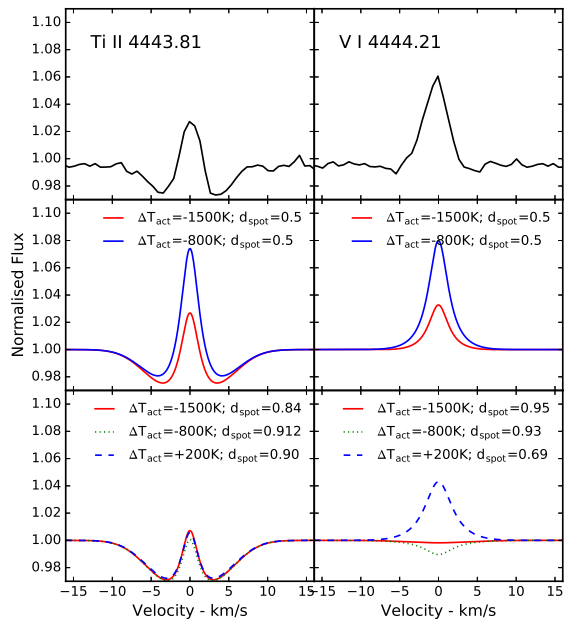}
	\vspace{-0.5cm}\caption{\textit{Top panels:} Two examples of narrow pseudo-emission profiles in
	the observed relative spectra from closely spaced ($\sim$0.4~\AA) and approximately
	equal strength lines.
	Both of these profiles show distinctly different morphologies, with the \ion{Ti}{ii}
	4443.8~\AA~line (left) displaying a clear pseudo-absorption trough.
	\textit{middle panels:} We show results for a simple model with a 4$\%$ spot feature
	at disk centre, having a continuum contrast corresponding to
	a $\Delta$T of $\sim$1400K and $\sim$800K.
	The line absorption in the spot is $\sim$50$\%$ weaker than
	the immaculate photosphere.
	For the left panel we have assumed that the immaculate line profile in
	the active spectrum is broadened by 2.5$\%$ relative to the inactive
	immaculate photosphere case -- this results in the observed absorption trough.
	In the right-hand panel, the line-widths in the active and inactive immaculate spectra are identical.
	\textit{bottom panels:} Models using the normalised
	line depth from VALD of the given temperature differences.
	An additional model of $\Delta$T= +200K is also show,
	which is a proxy for a plage rather than a spotted region.}
	\label{fig:model}
\end{figure}

The narrow `pseudo-emission' peaks also show differing profile shapes,
as demonstrated in Fig. \ref{fig:model} showing two closely separated lines,
\ion{Ti}{ii} 4443.81~\AA~and \ion{V}{i} 4444.21~\AA. For \ion{Ti}{ii} 4443.81~\AA,
we see a distinctive pseudo-absorption trough surrounding the emission peak,
which is not present in the neighbouring \ion{V}{i} 4444.21~\AA~line.
The stark difference between these two close-by lines also rules
out instrumental effects, which would not change so dramatically over
this short wavelength range.
We have attempted to simulate these two relative line shapes using a
simple model that consists of a limb-darkened disk representing the
`immaculate photosphere' and a circular patch at the
centre of the stellar disk representing a spotted region.
The spotted region covered 4$\%$ of the visible modelled stellar surface.
We generate Gaussian-shaped line profiles for each point on the star
assuming solid body rotation.
The line properties (e.g. depth and width) in each region are
free parameters in our model.
For the immaculate photosphere region we chose values that best
recreated the spectral lines of \ion{Ti}{ii} 4443.81~\AA~
and \ion{V}{i} 4444.21~\AA~as seen in the master low-activity template.
The depth and width of the line was set to 90$\%$ of the continuum
and 7.5 km s$^{-1}$, respectively.
For the spotted region we changed the continuum levels to $10\%$
and $30\%$ that of the immaculate photosphere which, at the
wavelength of the lines, represent a $\Delta$T of $\sim$1400K
and $\sim$800K cooler, respectively.
The line profiles in the spotted and immaculate photosphere regions 
were summed together to produce our high-activity line profile models,
a second model disk without a spotted region was generated
to model the low-activity line profile.
The high-activity line profile was divided by the low-activity to
produce the final relative line profile model for comparison to our
data.

As mentioned previously, the \ion{Ti}{ii} 4443.81~\AA~line
(left plots of Fig. \ref{fig:model}) displays a distinctive pseudo-absorption trough.
We found that to reproduce this feature changing the parameters of the
line profile in the spotted region was not enough.
\cite{Cegla2013} show that magnetic bright points (MBPs)
can be a source of line broadening (see, for example, their Fig. 2),
these MBPs exist across the whole surface of the star and are not
just constrained to spotted regions.
In our model we broaden the immaculate photosphere
(i.e. the non-spotted region)
of the active
spectrum by 2.5$\%$ relative to the inactive immaculate photosphere,
this represents the presence of more MBPs across the surface of \acb~
during its more active state.
The addition of this term, as well as weakening the absorption strength
of the line in the spotted region by $\sim$50$\%$ relative to the immaculate
photosphere, allows us to more accurately recover the feature
(the middle left panel of Fig. \ref{fig:model}).
For the \ion{V}{i} 4444.21~\AA~line (right plots of Fig. \ref{fig:model})
we did not need to invoke any broadening of the immaculate photosphere
to reproduce the feature (shown in the middle right of Fig. \ref{fig:model}).
We then re-ran this model using the \ion{Ti}{ii} 4443.81~and
\ion{V}{i} 4444.21~\AA~lines depths given by VALD \citep{Ryabchikova2015} for spot
temperatures with $\Delta$T of $\sim$1400K and $\sim$800K.
These models (bottom panels, Fig. \ref{fig:model}) show that, while we can
still reproduce the overall morphology of the \ion{Ti}{ii} 4443.81~line,
we cannot reproduce the \ion{V}{i} 4444.21~\AA~line assuming
a simple cool spot model.
To explore this further, we generated a model assuming
the spotted region was 200K hotter that the immaculate
photosphere as a proxy for a hotter plage region.
This gives a qualitatively better fit to the observed \ion{V}{i} 4444.21~\AA~line
feature, and demonstrates the complexity of modelling these lines.

The differing morphologies also imply a difference in the physical
processes that affect the line strength during changes in activity,
with some lines being more affected by magnetic activity that others.
We believe that using such lines in high precision RV measurements
could increase RV noise, which could be mitigated by looking only
at lines with weak (or no) sensitivity to stellar activity.

\vspace{-0.4cm}\subsection{Iron Lines Pseudo-Equivalent Width} \label{linewidths_section}
\begin{figure}
	\vspace{-1.0cm}\includegraphics[width=\columnwidth]{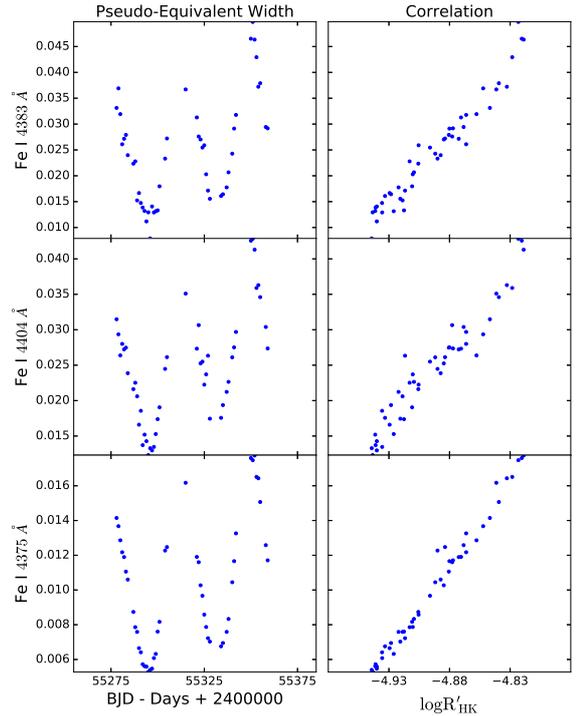}	
	\vspace{-0.6cm}\caption{\textit{left:} The pseudo-equivalent widths of the three lines \fei, \feii, \fei~
	plotted against day. Shown is the data during the March-June 2010
	period showing a clear periodic modulation.
	\textit{right:} Correlation between \logr~and the same 3 lines.}
	\label{fig:correlation}
\end{figure}

The peaks in the relative spectra shown in Fig. \ref{fig:temp_evolve} display
a correlation with \logr.
We calculate the pseudo-equivalent width of three close-by features
in the relative spectra: \fei, \feii~and \feiii in order to better trace the
changing strength of the lines.
The pseudo-equivalent width was taken as the area of a Gaussian
fit to each of the features.
The \feiii~line is a narrow feature that does not show any
pseudo-absorption trough, while the other two show broad features in the relative spectrum.
In Fig. \ref{fig:correlation} we show how the strengths of the three pseudo-equivalent widths
respond to changing levels of activity (as indicated by \logr) for the March-June 2010 period.
This period is characterised by a rotational modulation of the \logr, and
a similar variation is observed in all three lines.
This adds validity to the argument that the features are real, as producing
such a correlation by data processing or instrumental effects would be very difficult.

For the three lines, a clear periodic modulation of their pseudo-equivalent width
matches the rotational period of \acb.
The relative strengths of each of the \ion{Fe}{i} lines are plotted against
night (left panels of Fig. \ref{fig:correlation}), this suggests the changing strength of the
features is due to active regions rotating across the surface of the star.
The correlation of the pseudo-equivalent widths against \logr~is also shown in
the right panels of \ref{fig:correlation}, with all pseudo-equivalent widths
having a Pearson R value of $>0.96$.

This result more rigorously demonstrates the changes observed in
Fig. \ref{fig:temp_evolve} and shows the changing strength of both the 
broad and narrow features mimic the rotational variation seen in the value of \logr.

\vspace{-0.4cm}\subsection{Radial Velocity Variations}\label{lab:RVs}
If the features are activity driven, the position of the peak centre should vary
over the course of the rotation period.
To test this we selected  the first night in the March-June 2010 period and fit
Gaussians to all peaks in the relative spectrum that were stronger than $5\%$.
A template was generated from this fit that was then cross-correlated with
each of the relative spectra in the March-June 2010 period.

\begin{figure}
	\vspace{-0.5cm}\includegraphics[width=\columnwidth]{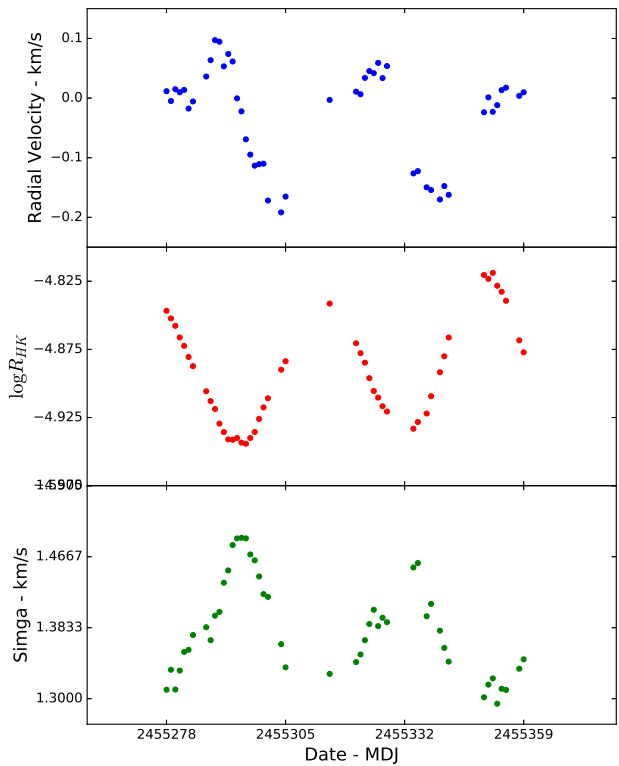}
	\vspace{-0.5cm}\caption{\textit{Top:} The radial velocity measurements,
	as defined by the peak of the cross corelation function, for the narrow
	features in the relative spectra. 
	\textit{Middle:} The \logr~as taken from \protect\cite{Dumusque2012}.
	\textit{Bottom:} The width of the cross correlation function.
	Both Top and Bottom plots show, with varying degrees of phase lag,
	the same period variation seen in \logr.}
	\label{fig:RV}
\end{figure}

We measured the velocity change of the relative peaks and found that the
lines show a peak-to-peak velocity change of $\sim$300 m s$^{-1}$
(top panel in Fig. \ref{fig:RV}).
Compared to  \logr~(middle panel of Fig. \ref{fig:RV}) the RVs of the
pseudo-emission peaks show a phase difference of approximately 90$^{\circ}$.
This can be understood by considering a rotating active region.
At disc centre, \logr~is at maximum as the active region has maximum
visibility, however the RV of the active region will be at 0 km s$^{-1}$
(relative to the systemic velocity of the system).
As the feature rotates out of view towards the stellar limb, foreshortening
will decrease \logr~, and the feature will become progressively red-shifted.
Conversely, the opposite trend occurs as the feature rotates back into view.
This is akin to the motion of apparent emission bumps due to spots through
stellar line-profiles, typically associated with more rapidly rotating stars
suitable for Doppler imaging \citep{Cameron2002, Vogt1983}. 

The width of the cross-correlation functions (CCFs) are shown on the bottom panel of Fig. \ref{fig:RV}.
These are anti-correlated with \logr, with the CCFs broadest when \logr~is lowest. This suggests
that active regions are more homogeneously distributed across the stellar surface during
times when the main active region has rotated out of direct view. The narrow CCF widths at high
\logr~then lend support for the presence of a localised highly active region (and hence spanning a limited range in stellar surface velocities). Of course, this is likely to be a simplistic picture due to the
probable presence of several active regions. The effects of plage and MBPs would also need to be considered as they
have been shown earlier in section \ref{AandD} to cause pseudo-emission
peaks of very different morphologies and may need to be considered separately in
our cross-correlation analyses.
As such, more in depth analysis of the effects of plage on relative peak
morphology and spots on the temperature sensitivity of the broad relative
peaks of the the \ion{Fe}{i} species is needed to better constrain the effects reported here.

\section{Conclusions}
We investigate the effects that stellar activity has on the spectrum of \acb.
We present evidence that the strength of a large number of spectral lines
changes due to the effects of active regions rotating in and out of view.
Relative spectra - created by taking high-activity spectra and dividing them by a
master low-activity template - show distinct narrow and broad features. These features have strengths and radial velocities that are modulated on the rotation period of the star, and
we show that they are associated with stellar activity.

The narrow peaks show differing morphologies, the most prominent
being pseudo-emission lines superimposed on top of broader absorption troughs.
We demonstrate that this can be explained if  absorption
lines from the `active immaculate' photosphere are broader than
their `inactive immaculate' photosphere counterparts.
We suggest that this could arise due to a higher filling factor of magnetic
bright points during the active phases, which may lead to a general
enhanced line-broadening across the stellar photosphere relative to the
inactive case.
In addition, broad features seen at \fei~and \feii~in the relative spectra belong to
temperature sensitive \ion{Fe}{i} lines, suggesting a change in the surface
temperature of \acb~(leading to an apparent change in spectral type).

The work presented here demonstrates the need to better understand the nature
and effects that activity and rotating active regions can have on the measurement
of spectral lines. In particular, the evidence that some strong lines can show distinct changes
between the active and inactive states, while other similarly strong lines do not show
such pronounced differences, suggests the ability to pre-select well-behaved lines suitable for
high RV-precision work. This will be explored in more detail in a further paper.
As we move closer to the launch of missions dedicated to the discovery of Earth-analog planets  (such as PLATO), such work may become critical in the RV confirmation of small, terrestrial
planets. In order to provide line-lists of stellar-activity insensitive
lines as a function of spectral type, we would urge the community to begin monitoring 
stellar activity cycles of likely mission targets such that bespoke low-activity comparison spectra may be obtained.

\vspace{-0.5cm}\section{acknowledgements}
AT acknowledges funding from DE and EdM would like to acknowledge the support
of the Michael West Fellowship. CAW acknowledges support from the STFC grant
ST/L000709/1.
This work has made use of the VALD database, operated at Uppsala
University, the Institute of Astronomy RAS in Moscow,
and the University of Vienna.
We would like to thank the anonymous referee for their
useful comments on this letter.

\vspace{-0.2cm}\bibliography{Reference}

\begin{thebibliography}{}
\makeatletter
\relax
\def\mn@urlcharsother{\let\do\@makeother \do\$\do\&\do\#\do\^\do\_\do\%\do\~}
\def\mn@doi{\begingroup\mn@urlcharsother \@ifnextchar [ {\mn@doi@}
  {\mn@doi@[]}}
\def\mn@doi@[#1]#2{\def\@tempa{#1}\ifx\@tempa\@empty \href
  {http://dx.doi.org/#2} {doi:#2}\else \href {http://dx.doi.org/#2} {#1}\fi
  \endgroup}
\def\mn@eprint#1#2{\mn@eprint@#1:#2::\@nil}
\def\mn@eprint@arXiv#1{\href {http://arxiv.org/abs/#1} {{\tt arXiv:#1}}}
\def\mn@eprint@dblp#1{\href {http://dblp.uni-trier.de/rec/bibtex/#1.xml}
  {dblp:#1}}
\def\mn@eprint@#1:#2:#3:#4\@nil{\def\@tempa {#1}\def\@tempb {#2}\def\@tempc
  {#3}\ifx \@tempc \@empty \let \@tempc \@tempb \let \@tempb \@tempa \fi \ifx
  \@tempb \@empty \def\@tempb {arXiv}\fi \@ifundefined
  {mn@eprint@\@tempb}{\@tempb:\@tempc}{\expandafter \expandafter \csname
  mn@eprint@\@tempb\endcsname \expandafter{\@tempc}}}

\bibitem[\protect\citeauthoryear{{Basri}, {Wilcots}  \& {Stout}}{{Basri}
  et~al.}{1989}]{Basri1989}
{Basri} G.,  {Wilcots} E.,   {Stout} N.,  1989, \mn@doi [\pasp]
  {10.1086/132464}, \href {http://adsabs.harvard.edu/abs/1989PASP..101..528B}
  {101, 528}

\bibitem[\protect\citeauthoryear{{Cegla}, {Shelyag}, {Watson}  \&
  {Mathioudakis}}{{Cegla} et~al.}{2013}]{Cegla2013}
{Cegla} H.~M.,  {Shelyag} S.,  {Watson} C.~A.,   {Mathioudakis} M.,  2013,
  \mn@doi [\apj] {10.1088/0004-637X/763/2/95}, \href
  {http://adsabs.harvard.edu/abs/2013ApJ...763...95C} {763, 95}

\bibitem[\protect\citeauthoryear{{Collier Cameron} \& {Donati}}{{Collier
  Cameron} \& {Donati}}{2002}]{Cameron2002}
{Collier Cameron} A.,  {Donati} J.-F.,  2002, \mn@doi [\mnras]
  {10.1046/j.1365-8711.2002.05147.x}, \href
  {http://adsabs.harvard.edu/abs/2002MNRAS.329L..23C} {329, L23}

\bibitem[\protect\citeauthoryear{{DeWarf}, {Datin}  \& {Guinan}}{{DeWarf}
  et~al.}{2010}]{DeWarf2010}
{DeWarf} L.~E.,  {Datin} K.~M.,   {Guinan} E.~F.,  2010, \mn@doi [\apj]
  {10.1088/0004-637X/722/1/343}, \href
  {http://adsabs.harvard.edu/abs/2010ApJ...722..343D} {722, 343}

\bibitem[\protect\citeauthoryear{{Dumusque} et~al.,}{{Dumusque}
  et~al.}{2012}]{Dumusque2012}
{Dumusque} X.,  et~al., 2012, \mn@doi [\nat] {10.1038/nature11572}, \href
  {http://adsabs.harvard.edu/abs/2012Natur.491..207D} {491, 207}

\bibitem[\protect\citeauthoryear{{Flores}, {Gonz{\'a}lez}, {Jaque Arancibia},
  {Buccino}  \& {Saffe}}{{Flores} et~al.}{2016}]{Flores2016}
{Flores} M.,  {Gonz{\'a}lez} J.~F.,  {Jaque Arancibia} M.,  {Buccino} A.,
  {Saffe} C.,  2016, \mn@doi [\aap] {10.1051/0004-6361/201628145}, \href
  {http://adsabs.harvard.edu/abs/2016A%26A...589A.135F} {589, A135}

\bibitem[\protect\citeauthoryear{{Giridhar}}{{Giridhar}}{2010}]{Giridhar2010}
{Giridhar} S.,  2010, Bulletin of the Astronomical Society of India, \href
  {http://adsabs.harvard.edu/abs/2010BASI...38....1G} {38, 1}

\bibitem[\protect\citeauthoryear{Hall, Lockwood  \& Skiff}{Hall
  et~al.}{2007}]{Hall2007}
Hall J.~C.,  Lockwood G.~W.,   Skiff B.~A.,  2007, The Astronomical Journal,
  133, 862

\bibitem[\protect\citeauthoryear{{Lovis} et~al.,}{{Lovis}
  et~al.}{2011}]{Lovis2011}
{Lovis} C.,  et~al., 2011, preprint, \href
  {http://adsabs.harvard.edu/abs/2011arXiv1107.5325L} {} (\mn@eprint {arXiv}
  {1107.5325})

\bibitem[\protect\citeauthoryear{{Ryabchikova}, {Piskunov}, {Kurucz},
  {Stempels}, {Heiter}, {Pakhomov}  \& {Barklem}}{{Ryabchikova}
  et~al.}{2015}]{Ryabchikova2015}
{Ryabchikova} T.,  {Piskunov} N.,  {Kurucz} R.~L.,  {Stempels} H.~C.,  {Heiter}
  U.,  {Pakhomov} Y.,   {Barklem} P.~S.,  2015, \mn@doi [\physscr]
  {10.1088/0031-8949/90/5/054005}, \href
  {http://adsabs.harvard.edu/abs/2015PhyS...90e4005R} {90, 054005}

\bibitem[\protect\citeauthoryear{{Vogt} \& {Penrod}}{{Vogt} \&
  {Penrod}}{1983}]{Vogt1983}
{Vogt} S.~S.,  {Penrod} G.~D.,  1983, \mn@doi [\pasp] {10.1086/131208}, \href
  {http://adsabs.harvard.edu/abs/1983PASP...95..565V} {95, 565}

\bibitem[\protect\citeauthoryear{{Wilson}}{{Wilson}}{1978}]{Wilson1978}
{Wilson} O.~C.,  1978, \mn@doi [\apj] {10.1086/156618}, \href
  {http://adsabs.harvard.edu/abs/1978ApJ...226..379W} {226, 379}

\makeatother
\end{thebibliography}
\bibliographystyle{mnras}

\bsp	
\label{lastpage}
\end{document}